\documentclass[prl,aps,twocolumn,showpacs,10pt,superscriptaddress]{revtex4-2}

\usepackage{graphicx}  
\graphicspath{{./Figures/}}
\usepackage{dcolumn}  
\usepackage[nooneline,hang]{subfigure}
\usepackage{amssymb, amsmath}

\usepackage{siunitx}
\DeclareSIUnit\torr{Torr}
\DeclareSIUnit\amagat{amg}

\usepackage{hyperref}

\newcommand{\PRLsection}[1]{\noindent \textit{#1} ---}

\usepackage{graphicx}
\usepackage{bm}
\usepackage[mathlines]{lineno}

\usepackage{color}
\definecolor{mygreen}{rgb}{0,0.5,0}
\definecolor{mygrey}{rgb}{0.5,0.5,0.5}
\definecolor{myred}{rgb}{0.75,0,0}
\definecolor{myblue}{rgb}{0,0,0.75}
\definecolor{mymagenta}{cmyk}{0,1,0,0.12}
\definecolor{mycyan}{cmyk}{1,0,0,0.12}
\definecolor{myorange}{rgb}{1.,0.5,0}
\definecolor{myviolet}{rgb}{0.6,0.15,0.6}
\definecolor{mybrown}{cmyk}{0,0.50,1,0.41}

\usepackage{ulem}

\newcommand{\supzero}{^{(0)}}
\newcommand{\supone}{^{(1)}}

\newcommand{\be}{\begin{equation}}
\newcommand{\ee}{\end{equation}}
\newcommand{\bea}{\begin{eqnarray}}
\newcommand{\eea}{\end{eqnarray}}

\newcommand{\bB}{{\bf B}}

\newcommand{\bF}{{\bf F}}

\newcommand{\Sph}{\mathcal{S}_{v}^{\mathrm{PSN}}}
\newcommand{\Sat}{\mathcal{S}_{v}^{\mathrm{SPN}}}
\newcommand{\Smba}{\mathcal{S}_{v}^\mathrm{MBA}}

\newcommand{\SBph}{\mathcal{S}_{B}^{\mathrm{PSN}}}
\newcommand{\SBat}{\mathcal{S}_{B}^{\mathrm{SPN}}}
\newcommand{\SBmba}{\mathcal{S}_{B}^\mathrm{MBA}}

\newcommand{\ICFO}{ICFO - Institut de Ci\`encies Fot\`oniques, The Barcelona Institute of Science and Technology, 08860 Castelldefels (Barcelona), Spain}
\newcommand{\ICREA}{ICREA - Instituci\'{o} Catalana de Recerca i Estudis Avan{\c{c}}ats, 08010 Barcelona, Spain}
\newcommand{\BARI}{Dipartimento Interateneo di Fisica, Universit\'{a}
degli Studi di Bari Aldo Moro, 70126 Bari, Italy}

\begin{document}

\newcommand{\thetitle}{ Squeezed light Bell Bloom OPM with varying atomic density vapor  }

\renewcommand{\thetitle}{
Quantum-enhanced magnetometry at optimal number density
}

\title{\thetitle}

\author{Charikleia Troullinou}
\email[Corresponding author: ]{charikleia.troullinou@icfo.eu}
\affiliation{\ICFO}
\author{Vito Giovanni  Lucivero}
\affiliation{\ICFO}
\affiliation{\BARI}
\author{Morgan W. Mitchell}
\email[Corresponding author: ]{morgan.mitchell@icfo.eu}
\affiliation{\ICFO}
\affiliation{\ICREA}

\date{\today}

\begin{abstract}
We study the use of squeezed probe light and evasion of measurement back-action to enhance the sensitivity and measurement bandwidth of an optically-pumped magnetometer (OPM) at sensitivity-optimal atom number density.
By experimental observation, and in agreement with quantum noise modeling, a spin-exchange-limited OPM probed with off-resonance laser light is shown to have an optimal sensitivity determined by density-dependent quantum noise contributions. Application of squeezed probe light boosts the OPM sensitivity beyond this laser-light optimum, allowing the OPM to achieve sensitivities that it cannot reach with coherent-state probing at any density. The observed quantum sensitivity enhancement at optimal number density is enabled by measurement back-action evasion.

\end{abstract}

\maketitle

Quantum sensitivity enhancement, i.e., the use of non-classical correlations to improve measurement sensitivity \cite{Helstrom1969, BraginskyBook1995, CavesPRD1981} is a topic of fundamental interest with practical implications for quantum-noise-limited optical interferometers \cite{LIGONP2011, AcernesePRL2019Short, TsePRL2019Short}, atomic clocks \cite{Schulte2020,BenedictoPRL2022}, magnetometers \cite{WolfgrammPRL2010, Vasilakis2015, MartinPRL2017}, and searches for physics beyond the Standard Model \cite{BackesN2021}.  The topic has been extensively studied in the particle-number-limited scenario, where paradigmatic models show power-law scaling of sensitivity with the available particle number for reviews see \cite{DegenRMP2017, PezzeRMP2018}. In contrast, many important instruments for contemporary applications \cite{Mei2020,Kitching2018,Aslam2023}, including optical interferometers \cite{Tse2019}, vapor- and gas-phase atomic clocks \cite{Knappe2004Clock,Newman2019}, magnetometers  \cite{Jimenez-Martinez2012, Lucivero2022} and gyroscopes \cite{KornackPRL2005}, employ abundant particles and are free to operate at a sensitivity-optimal particle number. The existence of such an optimum is incompatible with power-law sensitivity and implies mutual interaction of the sensing particles \cite{MitchellQST2017}, significant departures from paradigmatic models. The utility of non-classical states in this number-unconstrained scenario is a subtle question that we address here.


Atomic vapor sensors typically show an optimum quantum-limited sensitivity with vapor number density $n$ (and thus with particle number) due to a competition of factors: signal strength grows with increasing $n$, but may saturate due to density-dependent effects such as collisional line-broadening \cite{Lucivero2016}. Meanwhile, intrinsic noise sources including collisions and measurement back-action (MBA, disturbance of the atomic spins due to its optical readout) tend to increase with increasing $n$, creating the conditions for an $n$-optimum \cite{Lucivero2017}. Squeezing plays a dual role in this scenario: it reduces noise associated with the readout \textit{per se}, but the associated anti-squeezing can also increase MBA noise. 
It is thus not obvious that probe squeezing is beneficial in the number-optimized scenario.

Prior work on this topic includes theoretical study of spectroscopy of saturable media \cite{MitchellQST2017} and experimental spin-noise spectroscopy \cite{Lucivero2016, Lucivero2017}.  In magnetometry, Li et al.~\cite{LiJOSAB2022} studied a low-density magnetometer with resonant squeezed-light probing and modulated optical pumping, and observed an $n$-optimized signal-to-noise ratio (SNR) that improved with squeezing, with optimal sensitivity $\approx \SI{250}{\pico\tesla\per\sqrt\hertz}$. Using a similar system, Horrom et al.~\cite{HorromPRA2012} observed an $n$-optimized sensitivity of $\approx \SI{2}{\pico\tesla\per\sqrt\hertz}$, about two orders of magnitude more sensitive than Li et al. In this case, however, squeezed light improved the OPM sensitivity only at sub-optimal densities and left the $n$-optimal sensitivity unchanged to within experimental uncertainties. In both cases, the relatively low optimal $n$ was attributed to nonlinear optical noise \cite{NovikovaPRA2015} driven by the on-resonance probe.


\begin{figure*}[t]
    \centering
    \includegraphics[width=\textwidth]{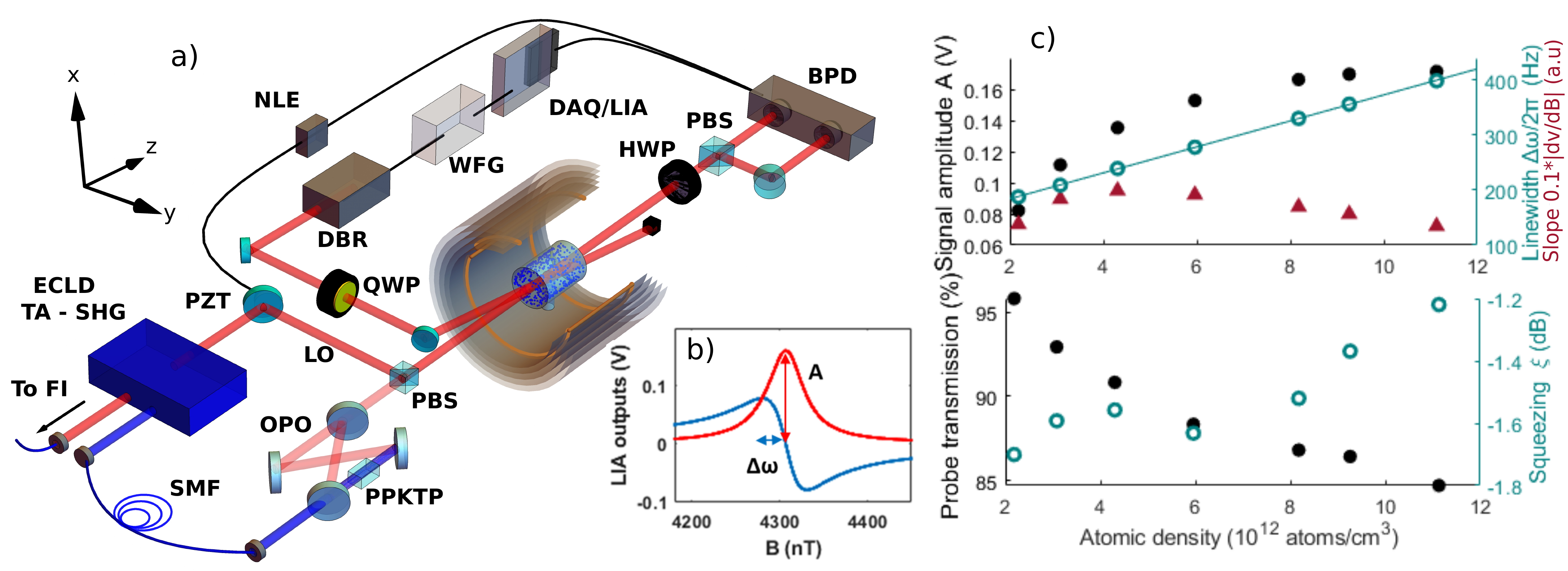}
\caption{\textbf{(a) Squeezed-light OPM experimental setup.} ECDL - Extended cavity diode laser, TA-SHG - tapered amplified second harmonic generation, FI - fiber interferometer, SMF - single mode fiber, OPO - optical parametric oscillator), PPKTP - nonlinear optical crystal, PZT - piezoelectric transducer, LO - local oscillator, PBS - polarizing beam splitter, QWP - quarter wave-plate, DBR - distributed Bragg reflector laser, WFG - waveform generator, BPD - balanced photo detection, HWP - half wave-plate, NLE - noise lock electronics, DAQ - data acquisition. A bias field along the x direction induces spin precession in the y-z plane. Optical pumping and probing are along $z$, resulting in a back-action evading geometry.  \textbf{(b) Polarization-rotation quadratures versus field strength.} Representative signal after demodulation by lock-in amplifier (LIA). \textbf{(c) Density dependence of signal parameters.} Signal amplitude, HWHM resonance linewidth with the linear fit,  and slope (top), probe transmission and degree of squeezing after atomic interaction (bottom) as a function of number density.  }
    \label{fig:experimentalSetup}
\end{figure*}

Here we demonstrate that a recently described quantum enhanced OPM \cite{Troullinou2021}, in addition to being about an order of magnitude more sensitive than Horrom et al., shows a significant quantum enhancement of the $n$-optimized sensitivity.  
To understand this behaviour, we extend the quantum noise model of Troullinou et al.~\cite{Troullinou2021} to describe the density dependence of the spin-projection noise (SPN), photon shot noise (PSN) and MBA contributions, with and without squeezing. The model is in good agreement with experiment and suggests that the improved $n$-optimized sensitivity is enabled by off-resonance probing, as in other high-sensitivity OPMs \cite{Smullin2009, Sheng2013, Perry2020, Gerginov2020, Lucivero2021}, and that the quantum advantage is enabled by MBA-evasion.


\PRLsection{Magnetometer operation}
As illustrated in Fig.~\ref{fig:experimentalSetup} (a), the OPM uses resonant Bell-Bloom \cite{BellBloom} optical pumping to produce a precessing spin polarization in a cell of \textsuperscript{87}Rb vapor in \SI{100}{\torr} of N\textsubscript{2} buffer gas. The precession is detected by paramagnetic Faraday rotation of a linearly polarized probe beam that overlaps with a frequency-modulated pump beam across the vapor cell. The optical rotation signal is demodulated with the optical pumping signal as a reference. The magnetometer signal is the \SI{90}{\degree} quadrature, proportional to $\Omega_L-\Omega$ for slowly varying magnetic fields, where $\Omega_L = \gamma B$ is the Larmor frequency, $\Omega/2\pi$ is the optical pumping modulation frequency and $\gamma$ the  bare \textsuperscript{87}Rb gyromagnetic ratio.  The OPM is operated with fixed pumping frequency $\Omega \approx 2\pi \times \SI{30}{\kilo\hertz}$ and with \textsuperscript{87}Rb number densities from \SI{2.18e12} to \SI{1.13e13}{atoms\per\centi\meter\cubed}. \footnote{The number density $n$ is adjusted by stabilizing the temperature $T$ of the ceramic oven enclosing the vapor cell. For \textsuperscript{87}Rb and 
$T>\SI{312.5}{\kelvin}$ we compute $n$ in \SI{}{atoms\per\centi\meter\cubed} using $\log_{10}(Tn) = 21.866 + A -B/T$  with coefficients $A = 4.312$ and $B=\SI{4040}{K}$
\cite{KongNC2020}.} 

A sub-threshold optical parametric oscillator (OPO) produces squeezed vacuum that is combined with laser output to produce polarization-squeezed probe light \cite{Predojevic2008}. 
The \SI{400}{\micro\watt} probe beam is blue detuned by \SI{20}{\giga\hertz} from the D$_1$ line, with optional squeezing of the $S_2$ Stokes component by \SI{2}{\decibel}, produced as detailed in \cite{Troullinou2021}. The OPO squeezer allows a probe detuning by multiple absorption linewidths, greatly reducing losses at high density.  
We observe probe transmission of \SI{95.8}{\percent} to \SI{84.7}{\percent} over the range of densities studied, and loss of squeezing consistent with simple absorption,  see Fig.~\ref{fig:experimentalSetup} (c)(bottom). 

To determine the magnetometer responsivity at fixed pumping modulation, the magnetic field $B$ is slowly scanned from \SI{4.2}{\micro\tesla} to \SI{4.45}{\micro\tesla}  around resonance $\mathrm{B}_0=\Omega/\gamma$, and the resulting polarimeter signal analyzed as in \cite{Troullinou2021}: the rotation signal is demodulated by a digital lock-in amplifier (LIA) to obtain the in-phase and \SI{90}{\degree} quadratures $u(B)$ and $v(B)$, respectively, shown in  Fig.~\ref{fig:experimentalSetup} (b). $u(B)$  is fit with a Lorentzian to obtain the 
signal amplitude $A \equiv u(B_0)$ and the half-width at half-maximum (HWHM) linewidth $\Delta\omega$ (expressed as an angular frequency using $\gamma = \SI{6.998}{\hertz\per\nano\tesla}$, as appropriate outside the SERF regime \cite{Savukov2005}). 
As shown in Fig.~\ref{fig:experimentalSetup} (c)(top), the resonance linewidth $\Delta \omega$ grows linearly with increasing $n$, due to collisional broadening, while $A$ shows saturating growth with increasing $n$ \cite{SupplementalMaterial}, resulting in a maximum of the slope ${dv}/{dB} =A / (2\gamma\Delta\omega)$ for a number density $n \approx \SI{4.3e12}{atoms\per\centi\meter\cubed}$.

\begin{figure}
    \centering
    \includegraphics[width = \columnwidth]{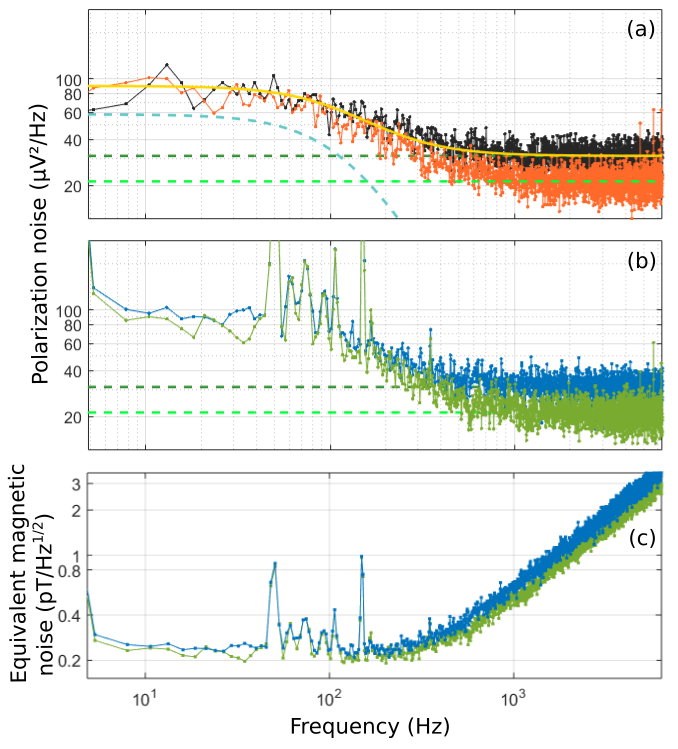}
        \caption{\textbf{Polarization rotation noise and equivalent magnetic noise spectra for T =\SI{95}{\celsius}} (a) The spin noise spectrum from unpolarized atoms with classical light (black) is fitted using Eq.~\ref{eq:SvInTermsOfSNAndSsigma} to obtain the standard quantum noise limit (dashed green) and the spin projection noise (dashed cyan line). By fitting the corresponding squeezed-light spectrum (orange), the photon shot noise is shown to be reduced. (light green dashed line). (b) Magnetometer spectral noise for polarized atoms. Compared with the corresponding noise spectra for unpolarized atoms the spectra show the same photon shot noise for classical and squeezed light probing. (c) Equivalent magnetic noise spectra for Bell-Bloom magnetometer probed with laser- (blue) and squeezed light (green). The data were acquired for vapor cell temperature $T = \SI{95}{\celsius}$, $n =\SI{4.3e12}{atoms\per\centi\meter\cubed}$, $ P_{\rm{probe}} = \SI{400}{\micro\watt}$, $P_{\rm{pump}} = \SI{500}{\micro\watt}$, $B_0 = \SI{4.3}{\micro\tesla}$. \SI{50}{\hertz} and \SI{150}{\hertz} peaks are power-line noise, peaks in the range \SI{60}{\hertz}--\SI{120}{\hertz} appear to be pump technical noise.
    }
    \label{fig:SNSandMAGspectra95C}
\end{figure}
\begin{figure*}[t]
	\centering
	\includegraphics[width=0.90\textwidth]{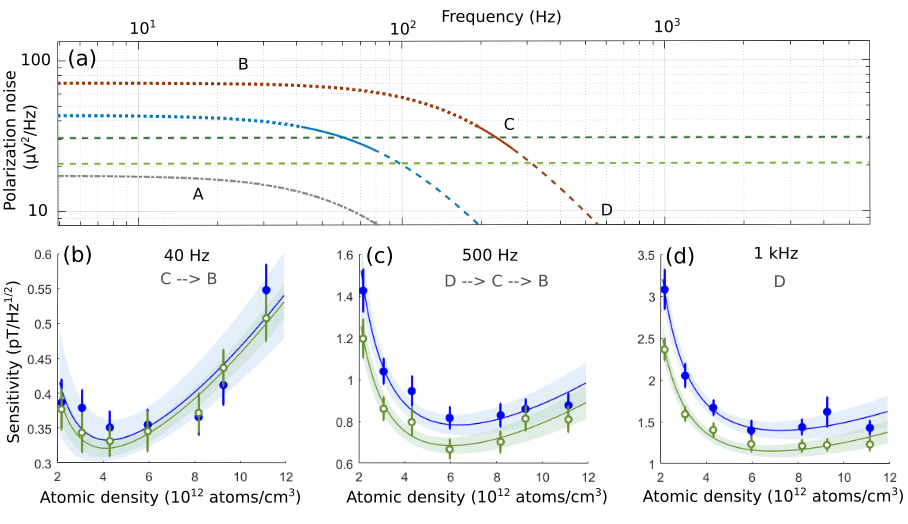}
	\caption{\textbf{Polarization noise and sensitivity for different atomic densities and analysis frequencies.}  (Top) The dark (light) green line shows the shot noise level with laser- (squeezed-light-) probing. The curves show the spin projection noise estimated from the fits of noise spectra for atomic densities of \SI{2.18e12}{atoms\per\centi\meter\cubed} (blue) and \SI{1.13e13}{atoms\per\centi\meter\cubed} (red) and extrapolated for \SI{0.5e12}{atoms\per\centi\meter\cubed} (grey). We distinguish four operational regimes indicated with different line texture A(dash-dotted): $\Sph > \Sat(\omega)\,\forall \omega $, B (dotted): $\Sph < \Sat(\omega)$, C (solid): $\Sph \sim \Sat(\omega)$ and  D (dashed): $\Sat(0)> \Sph >\Sat(\omega)$ for $\omega > \omega_{\SI{3}{\decibel}}$. (Bottom) The blue filled (green empty) circles and error bars indicate the mean and standard error of the mean of magnetometer sensitivity measured at a fixed analysis frequency for laser- (squeezed-) light probing. Solid lines and shaded regions show fits with Eq.~(\ref{eq:SBAgainWithMBAmain}) and \SI{95}{\percent} confidence intervals, respectively. At each analysis frequency, as the atomic density increases we observe a transition between the operational regimes.
 }
 \label{fig:DoubleSmiles}
\end{figure*}

At fixed magnetic field $B_0 = \SI{4.3}{\micro\tesla}$,  both laser- and squeezed light are used to probe the atomic ensemble with a range of densities and both in the unpumped condition, i.e., spin noise spectroscopy \cite{Lucivero2016}, and with Bell-Bloom optical pumping, i.e., as a magnetometer \cite{Troullinou2021}. The polarimeter signal is recorded for 50 acquisition cycles of \SI{0.5}{\second} each. For each acquisition cycle, we compute $\mathcal{S}_v(\omega)$, the single-sided power spectral density of $v$, by discrete Fourier transform with a Hann window. 
As calculated in \cite{SupplementalMaterial}, for the general case of a quantum noise limited magnetometer 
\bea
\label{eq:SvInTermsOfSNAndSsigma}
{\cal S}_v(\omega) = \Sph + {\cal L} (\omega)(\Sat + \Smba), 
\eea 
where $\Sph$, $\Sat$ and $\Smba$ are the contributions of the photon shot noise, spin projection noise and measurement back-action, respectively, where ${\cal L}(\omega) \equiv \Delta\omega^2/(\omega^2 + \Delta\omega^2)$ with \SI{3}{\decibel} response bandwidth $\Delta\omega$. 
 Apart from technical noise peaks, the magnetometer and spin projection noise spectra overlap for every atom number density studied. Each spin noise spectrum is fit using Eq.~\ref{eq:SvInTermsOfSNAndSsigma} with $\Smba = 0$ to account for the measurement back-action evasion. As seen in Fig.~\ref{fig:SNSandMAGspectra95C}, optical squeezing reduces the photon shot noise by the same amount for both the spin noise and magnetometry data, while leaving the spin projection noise intact. The top of Fig.~\ref{fig:DoubleSmiles} (a) shows ${\cal S}_v(\omega)$  from Eq.~\ref{eq:SvInTermsOfSNAndSsigma} with the observed means of $\Sph$ and $\Sat {\cal L} (\omega)$ for the lowest and highest density condition measured and an estimate for $\Sat$ at $n=\SI{0.5e12}{atoms\per\centi\meter\cubed}$, see \cite{SupplementalMaterial}.
The $\Sat$ contribution increases with increasing density, such that the cross-over from photon shot noise to spin projection noise shifts to higher frequency. As in \cite{Troullinou2021, Jimenez-Martinez2012, Gerginov2017}, we compute the equivalent magnetic noise density as $\mathcal{S}_B(\omega) = \mathcal{S}_v(\omega)|dv/dB|^{-2} [\mathcal{L}(\omega)]^{-1}$ using the measured $dv/dB$ and $\mathcal{L}(\omega)$ calculated using the values of $\Delta\omega$ from the data of Fig.~\ref{fig:experimentalSetup}(c)(top).  

\PRLsection{Quantum advantage in sensitivity}
The resulting sensitivities, as a function of number density and for representative signal frequencies, are shown in the bottom of Fig.~\ref{fig:DoubleSmiles}.  For low analysis frequencies, e.g. \SI{40}{\hertz} in Fig.~\ref{fig:DoubleSmiles}(b),  $\mathcal{S}_B$ shows a clear drop and then rise with increasing $n$.   For intermediate analysis frequencies, e.g. \SI{500}{\hertz} in Fig.~\ref{fig:DoubleSmiles} (c),  $\mathcal{S}_B$ shows a drop, plateau, and rise with increasing $n$.  For high analysis frequencies, e.g. \SI{1}{\kilo\hertz} in Fig.~\ref{fig:DoubleSmiles} (d),  $\mathcal{S}_B$ shows a drop and plateau, although the subsequent rise is not within the range of $n$ studied.

As described in \cite{SupplementalMaterial}, the  $S_B(\omega)$ derived in \cite{Troullinou2021} can be expressed in terms of number density $n$, the degree of squeezing $\xi^2$, and the probe and pump photon fluxes $\Phi_\mathrm{pr}$ and $\Phi_\mathrm{pump}$, respectively, as
\begin{eqnarray}
\label{eq:SBAgainWithMBAmain}
 \mathcal{S}_B(\omega,n) 
&\propto& 
X \xi^2 \frac{\Delta \omega^2  (\omega^2 + \Delta\omega^2)}{n^2  \Phi_\mathrm{pump}^2 \Phi_\mathrm{pr} }
+
\frac{\Delta \omega^3  }{n \Phi_\mathrm{pump}^2 } 
+
Y \frac{ \Phi_\mathrm{pr}}{\xi^2}
 , \hspace{6mm}
\end{eqnarray}
where $X$ and $Y$ are constants that depend on the probe detuning and degree of back-action evasion, respectively, but are  independent of $n$, $\Phi_\mathrm{pr}$ and $\Phi_\mathrm{pump}$. The terms are, in order, the PSN, SPN and MBA contributions $\SBph$, $\SBat$ and $\SBmba$. The magnetic resonance linewidth  is $\Delta \omega = \Gamma + n \Gamma_2 + \bar{P}$ where 
$n \Gamma_2$ is the collisional broadening and $\bar{P}$ is the pump-induced broadening. We note a few features of the model, i.e., Eq.~\ref{eq:SBAgainWithMBAmain}: with increasing $n$, $\SBmba$ is constant, $\SBat$ first falls as $n^{-1}$ before rising as $n^{2}$ when $\Gamma_2 n \gg \Gamma$, i.e., when collisional broadening dominates. $\SBph$ first falls as $n^{-2}$ before flattening out as $\SBph \propto n^{0}$ if $\omega \gg \Gamma$, then rising as $n^{2}$ when $\Gamma_2  n \gg \Gamma, \omega$. Polarization squeezing, i.e., $\xi^2 < 1$, is advantageous provided $\SBph(\omega) > \SBmba(\omega)$. These predictions agree with the observed effects of number density and squeezing in Fig.~\ref{fig:DoubleSmiles}(b-d).

We fit the measured $\mathcal{S}_B(\omega,n)$ with Eq.~(\ref{eq:SBAgainWithMBAmain}) for the case of measurement backaction evasion ($Y=0$), and with $\Gamma + \bar{P} = \SI{137}{\hertz}$, $ \Gamma_2 = \SI{23.6}{\hertz \centi\meter\cubed}$ from a linear fit to the $\Delta \omega$ data shown in Fig.~\ref{fig:experimentalSetup}(c). The resulting $\mathcal{S}_B(\omega,n)$ fits, shown in Fig.~\ref{fig:DoubleSmiles}(b-d), are in good agreement. 
We write the quantum advantage in number-optimized sensing as $\zeta(\nu) \equiv \min_n \mathcal{S}_B^{\mathrm{C}}(2 \pi \nu,n)/\min_n\mathcal{S}_B^{\mathrm{S}}(2 \pi \nu,n)$, where  the superscript $\mathrm{C}$ ($\mathrm{S}$) indicates coherent (squeezed) state probing. The fits shown in Fig.~\ref{fig:DoubleSmiles}(b-d) imply $\zeta(\nu) = \SI{0.31+-0.30}{\decibel}$, \SI{1.17+-0.26}{\decibel} and \SI{1.69+-0.31}{\decibel} for $\nu = \SI{40}{\hertz}$, $\SI{500}{\hertz}$ and $\SI{1}{\kilo\hertz}$, respectively. 
Optical squeezing thus significantly improves the $n$-optimized sensitivity at intermediate and high frequencies, and does not significantly worsen the sensitivity in the low-frequency, SPN-limited regime. This is a direct observation of evasion of measurement back-action, without which squeezing would add low-frequency noise in high density conditions. Squeezed-light probing improves also the \SI{3}{\decibel} measurement bandwidth 
$\omega_{\SI{3}{\decibel}}\equiv \Delta \omega \sqrt{ \Sat /\Sph +1}$ \cite{Troullinou2021}. At all measured densities $\omega_{\SI{3}{\decibel}}$ is observed to increase when squeezed light is used for probing,  see \cite{SupplementalMaterial}.

The observed quantum advantage is currently limited by the \SI{2}{\decibel} input probe squeezing, atomic absorption of the probe, which reduces output squeezing to $\approx \SI{1.6}{\decibel}$ at optimal density, and the SPN contribution. 
More squeezing, e.g.,~\SI{5.6}{\decibel} of atom-resonant squeezing in \cite{HanOE2016}, is possible with technical improvements. Absorption losses can be reduced at constant $\SBph$, $\SBat$ and $\SBmba$ by increasing  probe detuning while also increasing probe power, due to different scaling of the real and imaginary parts of the vector polarizability with optical detuning \cite{StocktonThesis2007}. SPN can be reduced by light narrowing \cite{Appelt1999} and spin-exchange-relaxation-free operation \cite{Savukov2005}.

\PRLsection{Conclusion} 
We have studied the effect of squeezed light on the sensitivity of an optically pumped magnetometer with a naturally-occurring optimal number density $n$, typical of vapor-based quantum sensors \cite{Jimenez-Martinez2012,ItoKobayashi2016,Lucivero2022}.  Squeezing is observed to improve the $n$-optimized sensitivity for the frequency regimes limited by PSN and jointly limited by PSN and SPN, without loss of sensitivity in the SPN-limited regime, and thus to achieve sensitivities that cannot be reached at any density using classical probe light. An analysis of the quantum noise dynamics shows that this number-unconstrained advantage is a consequence of measurement back-action evasion, a feature of spin-precession magnetometers with Faraday rotation probing \cite{ColangeloN2017,Troullinou2021}. Analogous number-unconstrained sensors that may similarly benefit include free-induction-decay OPMs \cite{Limes2020, ColangeloN2017}, atomic gyroscopes and co-magnetometers \cite{KornackPRL2005, Limes2018} and opto-mechanical sensors that employ dispersive optical probing to measure the angular momentum of nano-objects 
\cite{ZielinskaPRL2023b}.

\PRLsection{Acknowledgements}
We thank Sven Bodenstedt, Maria Hernandez Ruiz, Kostas Mouloudakis and Aleksandra Sierant for helpful readings of the manuscript. 
This work was supported by the European Commission project OPMMEG (101099379);  H2020 Marie Sk{\l}odowska-Curie Actions project PROBIST (Grant Agreement No. 754510) and Spanish Ministry of Science MCIN with funding from European Union NextGenerationEU (PRTR-C17.I1)  by Generalitat de Catalunya ``Severo Ochoa'' Center of Excellence CEX2019-000910-S; projects SAPONARIA (PID2021-123813NB-I00) and MARICHAS (PID2021-126059OA-I00) funded by MCIN/ AEI /10.13039/501100011033/ FEDER, EU; Generalitat de Catalunya through the CERCA program;  Ag\`{e}ncia de Gesti\'{o} d'Ajuts Universitaris i de Recerca Grants No. 2017-SGR-1354 and 2021-SGR-01453; Fundaci\'{o} Privada Cellex; Fundaci\'{o} Mir-Puig; V.G. Lucivero acknowledges financial support from European Union NextGenerationEU (PNRR MUR project PE0000023 -- NQSTI).
This project has received funding from the European Defence Fund (EDF) under grant agreement EDF-2021-DIS-RDIS-ADEQUADE.  Funded by the European Union. Views and opinions expressed are however those of the author(s) only and do not necessarily reflect those of the European Union or the European Commission. Neither the European Union nor the granting authority can be held responsible for them.

\bibliographystyle{apsrev4-2}
\bibliography{./biblio/DSBib}

\begin{thebibliography}{52}%
\makeatletter
\providecommand \@ifxundefined [1]{%
 \@ifx{#1\undefined}
}%
\providecommand \@ifnum [1]{%
 \ifnum #1\expandafter \@firstoftwo
 \else \expandafter \@secondoftwo
 \fi
}%
\providecommand \@ifx [1]{%
 \ifx #1\expandafter \@firstoftwo
 \else \expandafter \@secondoftwo
 \fi
}%
\providecommand \natexlab [1]{#1}%
\providecommand \enquote  [1]{``#1''}%
\providecommand \bibnamefont  [1]{#1}%
\providecommand \bibfnamefont [1]{#1}%
\providecommand \citenamefont [1]{#1}%
\providecommand \href@noop [0]{\@secondoftwo}%
\providecommand \href [0]{\begingroup \@sanitize@url \@href}%
\providecommand \@href[1]{\@@startlink{#1}\@@href}%
\providecommand \@@href[1]{\endgroup#1\@@endlink}%
\providecommand \@sanitize@url [0]{\catcode `\\12\catcode `\$12\catcode
  `\&12\catcode `\#12\catcode `\^12\catcode `\_12\catcode `\%12\relax}%
\providecommand \@@startlink[1]{}%
\providecommand \@@endlink[0]{}%
\providecommand \url  [0]{\begingroup\@sanitize@url \@url }%
\providecommand \@url [1]{\endgroup\@href {#1}{\urlprefix }}%
\providecommand \urlprefix  [0]{URL }%
\providecommand \Eprint [0]{\href }%
\providecommand \doibase [0]{https://doi.org/}%
\providecommand \selectlanguage [0]{\@gobble}%
\providecommand \bibinfo  [0]{\@secondoftwo}%
\providecommand \bibfield  [0]{\@secondoftwo}%
\providecommand \translation [1]{[#1]}%
\providecommand \BibitemOpen [0]{}%
\providecommand \bibitemStop [0]{}%
\providecommand \bibitemNoStop [0]{.\EOS\space}%
\providecommand \EOS [0]{\spacefactor3000\relax}%
\providecommand \BibitemShut  [1]{\csname bibitem#1\endcsname}%
\let\auto@bib@innerbib\@empty
\bibitem [{\citenamefont {Helstrom}(1969)}]{Helstrom1969}%
  \BibitemOpen
  \bibfield  {author} {\bibinfo {author} {\bibfnamefont {C.~W.}\ \bibnamefont
  {Helstrom}},\ }\href {https://doi.org/10.1007/BF01007479} {\bibfield
  {journal} {\bibinfo  {journal} {Journal of Statistical Physics}\ }\textbf
  {\bibinfo {volume} {1}},\ \bibinfo {pages} {231} (\bibinfo {year}
  {1969})}\BibitemShut {NoStop}%
\bibitem [{\citenamefont {Braginsky}\ \emph {et~al.}(1995)\citenamefont
  {Braginsky}, \citenamefont {Khalili},\ and\ \citenamefont
  {Thorne}}]{BraginskyBook1995}%
  \BibitemOpen
  \bibfield  {author} {\bibinfo {author} {\bibfnamefont {V.}~\bibnamefont
  {Braginsky}}, \bibinfo {author} {\bibfnamefont {F.}~\bibnamefont {Khalili}},\
  and\ \bibinfo {author} {\bibfnamefont {K.}~\bibnamefont {Thorne}},\ }\href
  {https://books.google.es/books?id=HHKRzXIVdxcC} {\emph {\bibinfo {title}
  {Quantum Measurement}}}\ (\bibinfo  {publisher} {Cambridge University
  Press},\ \bibinfo {year} {1995})\BibitemShut {NoStop}%
\bibitem [{\citenamefont {Caves}(1981)}]{CavesPRD1981}%
  \BibitemOpen
  \bibfield  {author} {\bibinfo {author} {\bibfnamefont {C.~M.}\ \bibnamefont
  {Caves}},\ }\href {https://doi.org/10.1103/PhysRevD.23.1693} {\bibfield
  {journal} {\bibinfo  {journal} {Phys. Rev. D}\ }\textbf {\bibinfo {volume}
  {23}},\ \bibinfo {pages} {1693} (\bibinfo {year} {1981})}\BibitemShut
  {NoStop}%
\bibitem [{\citenamefont {{The {LIGO} Scientific
  Collaboration}}(2011)}]{LIGONP2011}%
  \BibitemOpen
  \bibfield  {author} {\bibinfo {author} {\bibnamefont {{The {LIGO} Scientific
  Collaboration}}},\ }\href {http://dx.doi.org/10.1038/nphys2083} {\bibfield
  {journal} {\bibinfo  {journal} {Nat Phys}\ }\textbf {\bibinfo {volume} {7}},\
  \bibinfo {pages} {962} (\bibinfo {year} {2011})}\BibitemShut {NoStop}%
\bibitem [{\citenamefont {Acernese}\ and\ \citenamefont {{\textit et
  al.}}(2019)}]{AcernesePRL2019Short}%
  \BibitemOpen
  \bibfield  {author} {\bibinfo {author} {\bibfnamefont {F.}~\bibnamefont
  {Acernese}}\ and\ \bibinfo {author} {\bibnamefont {{\textit et al.}}},\
  }\href {https://doi.org/10.1103/PhysRevLett.123.231108} {\bibfield  {journal}
  {\bibinfo  {journal} {Phys. Rev. Lett.}\ }\textbf {\bibinfo {volume} {123}},\
  \bibinfo {pages} {231108} (\bibinfo {year} {2019})}\BibitemShut {NoStop}%
\bibitem [{\citenamefont {Tse}\ and\ \citenamefont {{\it et
  al.}}(2019)}]{TsePRL2019Short}%
  \BibitemOpen
  \bibfield  {author} {\bibinfo {author} {\bibfnamefont {M.}~\bibnamefont
  {Tse}}\ and\ \bibinfo {author} {\bibnamefont {{\it et al.}}},\ }\href
  {https://doi.org/10.1103/PhysRevLett.123.231107} {\bibfield  {journal}
  {\bibinfo  {journal} {Phys. Rev. Lett.}\ }\textbf {\bibinfo {volume} {123}},\
  \bibinfo {pages} {231107} (\bibinfo {year} {2019})}\BibitemShut {NoStop}%
\bibitem [{\citenamefont {Schulte}\ \emph {et~al.}(2020)\citenamefont
  {Schulte}, \citenamefont {Lisdat}, \citenamefont {Schmidt}, \citenamefont
  {Sterr},\ and\ \citenamefont {Hammerer}}]{Schulte2020}%
  \BibitemOpen
  \bibfield  {author} {\bibinfo {author} {\bibfnamefont {M.}~\bibnamefont
  {Schulte}}, \bibinfo {author} {\bibfnamefont {C.}~\bibnamefont {Lisdat}},
  \bibinfo {author} {\bibfnamefont {P.~O.}\ \bibnamefont {Schmidt}}, \bibinfo
  {author} {\bibfnamefont {U.}~\bibnamefont {Sterr}},\ and\ \bibinfo {author}
  {\bibfnamefont {K.}~\bibnamefont {Hammerer}},\ }\href
  {https://doi.org/10.1038/s41467-020-19403-7} {\bibfield  {journal} {\bibinfo
  {journal} {Nature Communications}\ }\textbf {\bibinfo {volume} {11}},\
  \bibinfo {pages} {5955} (\bibinfo {year} {2020})}\BibitemShut {NoStop}%
\bibitem [{\citenamefont {Orenes}\ \emph {et~al.}(2022)\citenamefont {Orenes},
  \citenamefont {Sewell}, \citenamefont {Lodewyck},\ and\ \citenamefont
  {Mitchell}}]{BenedictoPRL2022}%
  \BibitemOpen
  \bibfield  {author} {\bibinfo {author} {\bibfnamefont {D.~B.}\ \bibnamefont
  {Orenes}}, \bibinfo {author} {\bibfnamefont {R.~J.}\ \bibnamefont {Sewell}},
  \bibinfo {author} {\bibfnamefont {J.}~\bibnamefont {Lodewyck}},\ and\
  \bibinfo {author} {\bibfnamefont {M.~W.}\ \bibnamefont {Mitchell}},\ }\href
  {https://doi.org/10.1103/PhysRevLett.128.153201} {\bibfield  {journal}
  {\bibinfo  {journal} {Phys. Rev. Lett.}\ }\textbf {\bibinfo {volume} {128}},\
  \bibinfo {pages} {153201} (\bibinfo {year} {2022})}\BibitemShut {NoStop}%
\bibitem [{\citenamefont {Wolfgramm}\ \emph {et~al.}(2010)\citenamefont
  {Wolfgramm}, \citenamefont {Cer\`e}, \citenamefont {Beduini}, \citenamefont
  {Predojevi\ifmmode~\acute{c}\else \'{c}\fi{}}, \citenamefont {Koschorreck},\
  and\ \citenamefont {Mitchell}}]{WolfgrammPRL2010}%
  \BibitemOpen
  \bibfield  {author} {\bibinfo {author} {\bibfnamefont {F.}~\bibnamefont
  {Wolfgramm}}, \bibinfo {author} {\bibfnamefont {A.}~\bibnamefont {Cer\`e}},
  \bibinfo {author} {\bibfnamefont {F.~A.}\ \bibnamefont {Beduini}}, \bibinfo
  {author} {\bibfnamefont {A.}~\bibnamefont {Predojevi\ifmmode~\acute{c}\else
  \'{c}\fi{}}}, \bibinfo {author} {\bibfnamefont {M.}~\bibnamefont
  {Koschorreck}},\ and\ \bibinfo {author} {\bibfnamefont {M.~W.}\ \bibnamefont
  {Mitchell}},\ }\href {https://doi.org/10.1103/PhysRevLett.105.053601}
  {\bibfield  {journal} {\bibinfo  {journal} {Phys. Rev. Lett.}\ }\textbf
  {\bibinfo {volume} {105}},\ \bibinfo {pages} {053601} (\bibinfo {year}
  {2010})}\BibitemShut {NoStop}%
\bibitem [{\citenamefont {Vasilakis}\ \emph {et~al.}(2015)\citenamefont
  {Vasilakis}, \citenamefont {Shen}, \citenamefont {Jensen}, \citenamefont
  {Balabas}, \citenamefont {Salart}, \citenamefont {Chen},\ and\ \citenamefont
  {Polzik}}]{Vasilakis2015}%
  \BibitemOpen
  \bibfield  {author} {\bibinfo {author} {\bibfnamefont {G.}~\bibnamefont
  {Vasilakis}}, \bibinfo {author} {\bibfnamefont {H.}~\bibnamefont {Shen}},
  \bibinfo {author} {\bibfnamefont {K.}~\bibnamefont {Jensen}}, \bibinfo
  {author} {\bibfnamefont {M.}~\bibnamefont {Balabas}}, \bibinfo {author}
  {\bibfnamefont {D.}~\bibnamefont {Salart}}, \bibinfo {author} {\bibfnamefont
  {B.}~\bibnamefont {Chen}},\ and\ \bibinfo {author} {\bibfnamefont {E.~S.}\
  \bibnamefont {Polzik}},\ }\href {http://dx.doi.org/10.1038/nphys3280}
  {\bibfield  {journal} {\bibinfo  {journal} {Nature Physics}\ }\textbf
  {\bibinfo {volume} {11}},\ \bibinfo {pages} {389} (\bibinfo {year}
  {2015})}\BibitemShut {NoStop}%
\bibitem [{\citenamefont {Martin~Ciurana}\ \emph {et~al.}(2017)\citenamefont
  {Martin~Ciurana}, \citenamefont {Colangelo}, \citenamefont
  {Slodi\ifmmode~\check{c}\else \v{c}\fi{}ka}, \citenamefont {Sewell},\ and\
  \citenamefont {Mitchell}}]{MartinPRL2017}%
  \BibitemOpen
  \bibfield  {author} {\bibinfo {author} {\bibfnamefont {F.}~\bibnamefont
  {Martin~Ciurana}}, \bibinfo {author} {\bibfnamefont {G.}~\bibnamefont
  {Colangelo}}, \bibinfo {author} {\bibfnamefont {L.}~\bibnamefont
  {Slodi\ifmmode~\check{c}\else \v{c}\fi{}ka}}, \bibinfo {author}
  {\bibfnamefont {R.~J.}\ \bibnamefont {Sewell}},\ and\ \bibinfo {author}
  {\bibfnamefont {M.~W.}\ \bibnamefont {Mitchell}},\ }\href
  {https://link.aps.org/doi/10.1103/PhysRevLett.119.043603} {\bibfield
  {journal} {\bibinfo  {journal} {Phys. Rev. Lett.}\ }\textbf {\bibinfo
  {volume} {119}},\ \bibinfo {pages} {043603} (\bibinfo {year}
  {2017})}\BibitemShut {NoStop}%
\bibitem [{\citenamefont {Backes}\ \emph {et~al.}(2021)\citenamefont {Backes}
  \emph {et~al.}}]{BackesN2021}%
  \BibitemOpen
  \bibfield  {author} {\bibinfo {author} {\bibfnamefont {K.~M.}\ \bibnamefont
  {Backes}} \emph {et~al.},\ }\href
  {https://doi.org/10.1038/s41586-021-03226-7} {\bibfield  {journal} {\bibinfo
  {journal} {Nature}\ }\textbf {\bibinfo {volume} {590}},\ \bibinfo {pages}
  {238} (\bibinfo {year} {2021})}\BibitemShut {NoStop}%
\bibitem [{\citenamefont {Degen}\ \emph {et~al.}(2017)\citenamefont {Degen},
  \citenamefont {Reinhard},\ and\ \citenamefont {Cappellaro}}]{DegenRMP2017}%
  \BibitemOpen
  \bibfield  {author} {\bibinfo {author} {\bibfnamefont {C.~L.}\ \bibnamefont
  {Degen}}, \bibinfo {author} {\bibfnamefont {F.}~\bibnamefont {Reinhard}},\
  and\ \bibinfo {author} {\bibfnamefont {P.}~\bibnamefont {Cappellaro}},\
  }\href {https://doi.org/10.1103/RevModPhys.89.035002} {\bibfield  {journal}
  {\bibinfo  {journal} {Rev. Mod. Phys.}\ }\textbf {\bibinfo {volume} {89}},\
  \bibinfo {pages} {035002} (\bibinfo {year} {2017})}\BibitemShut {NoStop}%
\bibitem [{\citenamefont {Pezz\`e}\ \emph {et~al.}(2018)\citenamefont
  {Pezz\`e}, \citenamefont {Smerzi}, \citenamefont {Oberthaler}, \citenamefont
  {Schmied},\ and\ \citenamefont {Treutlein}}]{PezzeRMP2018}%
  \BibitemOpen
  \bibfield  {author} {\bibinfo {author} {\bibfnamefont {L.}~\bibnamefont
  {Pezz\`e}}, \bibinfo {author} {\bibfnamefont {A.}~\bibnamefont {Smerzi}},
  \bibinfo {author} {\bibfnamefont {M.~K.}\ \bibnamefont {Oberthaler}},
  \bibinfo {author} {\bibfnamefont {R.}~\bibnamefont {Schmied}},\ and\ \bibinfo
  {author} {\bibfnamefont {P.}~\bibnamefont {Treutlein}},\ }\href
  {https://doi.org/10.1103/RevModPhys.90.035005} {\bibfield  {journal}
  {\bibinfo  {journal} {Rev. Mod. Phys.}\ }\textbf {\bibinfo {volume} {90}},\
  \bibinfo {pages} {035005} (\bibinfo {year} {2018})}\BibitemShut {NoStop}%
\bibitem [{\citenamefont {Fu}\ \emph {et~al.}(2020)\citenamefont {Fu},
  \citenamefont {Iwata}, \citenamefont {Wickenbrock},\ and\ \citenamefont
  {Budker}}]{Mei2020}%
  \BibitemOpen
  \bibfield  {author} {\bibinfo {author} {\bibfnamefont {K.-M.~C.}\
  \bibnamefont {Fu}}, \bibinfo {author} {\bibfnamefont {G.~Z.}\ \bibnamefont
  {Iwata}}, \bibinfo {author} {\bibfnamefont {A.}~\bibnamefont {Wickenbrock}},\
  and\ \bibinfo {author} {\bibfnamefont {D.}~\bibnamefont {Budker}},\ }\href
  {https://doi.org/10.1116/5.0025186} {\bibfield  {journal} {\bibinfo
  {journal} {AVS Quantum Science}\ }\textbf {\bibinfo {volume} {2}},\ \bibinfo
  {pages} {044702} (\bibinfo {year} {2020})}\BibitemShut {NoStop}%
\bibitem [{\citenamefont {Kitching}(2018)}]{Kitching2018}%
  \BibitemOpen
  \bibfield  {author} {\bibinfo {author} {\bibfnamefont {J.}~\bibnamefont
  {Kitching}},\ }\href {https://doi.org/10.1063/1.5026238} {\bibfield
  {journal} {\bibinfo  {journal} {Applied Physics Reviews}\ }\textbf {\bibinfo
  {volume} {5}},\ \bibinfo {pages} {031302} (\bibinfo {year}
  {2018})}\BibitemShut {NoStop}%
\bibitem [{\citenamefont {Aslam}\ \emph {et~al.}(2023)\citenamefont {Aslam},
  \citenamefont {Zhou}, \citenamefont {Urbach}, \citenamefont {Turner},
  \citenamefont {Walsworth}, \citenamefont {Lukin},\ and\ \citenamefont
  {Park}}]{Aslam2023}%
  \BibitemOpen
  \bibfield  {author} {\bibinfo {author} {\bibfnamefont {N.}~\bibnamefont
  {Aslam}}, \bibinfo {author} {\bibfnamefont {H.}~\bibnamefont {Zhou}},
  \bibinfo {author} {\bibfnamefont {E.~K.}\ \bibnamefont {Urbach}}, \bibinfo
  {author} {\bibfnamefont {M.~J.}\ \bibnamefont {Turner}}, \bibinfo {author}
  {\bibfnamefont {R.~L.}\ \bibnamefont {Walsworth}}, \bibinfo {author}
  {\bibfnamefont {M.~D.}\ \bibnamefont {Lukin}},\ and\ \bibinfo {author}
  {\bibfnamefont {H.}~\bibnamefont {Park}},\ }\href
  {https://doi.org/10.1038/s42254-023-00558-3} {\bibfield  {journal} {\bibinfo
  {journal} {Nature Reviews Physics}\ }\textbf {\bibinfo {volume} {5}},\
  \bibinfo {pages} {157} (\bibinfo {year} {2023})}\BibitemShut {NoStop}%
\bibitem [{\citenamefont {Tse}\ \emph {et~al.}(2019)\citenamefont {Tse} \emph
  {et~al.}}]{Tse2019}%
  \BibitemOpen
  \bibfield  {author} {\bibinfo {author} {\bibfnamefont {M.}~\bibnamefont
  {Tse}} \emph {et~al.},\ }\href
  {https://doi.org/10.1103/PhysRevLett.123.231107} {\bibfield  {journal}
  {\bibinfo  {journal} {Phys. Rev. Lett.}\ }\textbf {\bibinfo {volume} {123}},\
  \bibinfo {pages} {231107} (\bibinfo {year} {2019})}\BibitemShut {NoStop}%
\bibitem [{\citenamefont {Knappe}\ \emph {et~al.}(2004)\citenamefont {Knappe},
  \citenamefont {Shah}, \citenamefont {Schwindt}, \citenamefont {Hollberg},
  \citenamefont {Kitching}, \citenamefont {Liew},\ and\ \citenamefont
  {Moreland}}]{Knappe2004Clock}%
  \BibitemOpen
  \bibfield  {author} {\bibinfo {author} {\bibfnamefont {S.}~\bibnamefont
  {Knappe}}, \bibinfo {author} {\bibfnamefont {V.}~\bibnamefont {Shah}},
  \bibinfo {author} {\bibfnamefont {P.~D.~D.}\ \bibnamefont {Schwindt}},
  \bibinfo {author} {\bibfnamefont {L.}~\bibnamefont {Hollberg}}, \bibinfo
  {author} {\bibfnamefont {J.}~\bibnamefont {Kitching}}, \bibinfo {author}
  {\bibfnamefont {L.-A.}\ \bibnamefont {Liew}},\ and\ \bibinfo {author}
  {\bibfnamefont {J.}~\bibnamefont {Moreland}},\ }\href
  {https://aip.scitation.org/doi/abs/10.1063/1.1787942} {\bibfield  {journal}
  {\bibinfo  {journal} {Applied Physics Letters}\ }\textbf {\bibinfo {volume}
  {85}},\ \bibinfo {pages} {1460} (\bibinfo {year} {2004})}\BibitemShut
  {NoStop}%
\bibitem [{\citenamefont {Newman}\ \emph {et~al.}(2019)\citenamefont {Newman}
  \emph {et~al.}}]{Newman2019}%
  \BibitemOpen
  \bibfield  {author} {\bibinfo {author} {\bibfnamefont {Z.~L.}\ \bibnamefont
  {Newman}} \emph {et~al.},\ }\href
  {https://opg.optica.org/optica/abstract.cfm?URI=optica-6-5-680} {\bibfield
  {journal} {\bibinfo  {journal} {Optica}\ }\textbf {\bibinfo {volume} {6}},\
  \bibinfo {pages} {680} (\bibinfo {year} {2019})}\BibitemShut {NoStop}%
\bibitem [{\citenamefont {Jim\'{e}nez-Mart\'{i}nez}\ \emph
  {et~al.}(2012)\citenamefont {Jim\'{e}nez-Mart\'{i}nez}, \citenamefont
  {Griffith}, \citenamefont {Knappe}, \citenamefont {Kitching},\ and\
  \citenamefont {Prouty}}]{Jimenez-Martinez2012}%
  \BibitemOpen
  \bibfield  {author} {\bibinfo {author} {\bibfnamefont {R.}~\bibnamefont
  {Jim\'{e}nez-Mart\'{i}nez}}, \bibinfo {author} {\bibfnamefont {W.~C.}\
  \bibnamefont {Griffith}}, \bibinfo {author} {\bibfnamefont {S.}~\bibnamefont
  {Knappe}}, \bibinfo {author} {\bibfnamefont {J.}~\bibnamefont {Kitching}},\
  and\ \bibinfo {author} {\bibfnamefont {M.}~\bibnamefont {Prouty}},\ }\href
  {https://doi.org/10.1364/JOSAB.29.003398} {\bibfield  {journal} {\bibinfo
  {journal} {J. Opt. Soc. Am. B}\ }\textbf {\bibinfo {volume} {29}},\ \bibinfo
  {pages} {3398} (\bibinfo {year} {2012})}\BibitemShut {NoStop}%
\bibitem [{\citenamefont {Lucivero}\ \emph {et~al.}(2022)\citenamefont
  {Lucivero}, \citenamefont {Lee}, \citenamefont {Limes}, \citenamefont
  {Foley}, \citenamefont {Kornack},\ and\ \citenamefont
  {Romalis}}]{Lucivero2022}%
  \BibitemOpen
  \bibfield  {author} {\bibinfo {author} {\bibfnamefont {V.}~\bibnamefont
  {Lucivero}}, \bibinfo {author} {\bibfnamefont {W.}~\bibnamefont {Lee}},
  \bibinfo {author} {\bibfnamefont {M.}~\bibnamefont {Limes}}, \bibinfo
  {author} {\bibfnamefont {E.}~\bibnamefont {Foley}}, \bibinfo {author}
  {\bibfnamefont {T.}~\bibnamefont {Kornack}},\ and\ \bibinfo {author}
  {\bibfnamefont {M.}~\bibnamefont {Romalis}},\ }\href
  {https://link.aps.org/doi/10.1103/PhysRevApplied.18.L021001} {\bibfield
  {journal} {\bibinfo  {journal} {Phys. Rev. Applied}\ }\textbf {\bibinfo
  {volume} {18}},\ \bibinfo {pages} {L021001} (\bibinfo {year}
  {2022})}\BibitemShut {NoStop}%
\bibitem [{\citenamefont {Kornack}\ \emph {et~al.}(2005)\citenamefont
  {Kornack}, \citenamefont {Ghosh},\ and\ \citenamefont
  {Romalis}}]{KornackPRL2005}%
  \BibitemOpen
  \bibfield  {author} {\bibinfo {author} {\bibfnamefont {T.~W.}\ \bibnamefont
  {Kornack}}, \bibinfo {author} {\bibfnamefont {R.~K.}\ \bibnamefont {Ghosh}},\
  and\ \bibinfo {author} {\bibfnamefont {M.~V.}\ \bibnamefont {Romalis}},\
  }\href {https://doi.org/10.1103/PhysRevLett.95.230801} {\bibfield  {journal}
  {\bibinfo  {journal} {Phys. Rev. Lett.}\ }\textbf {\bibinfo {volume} {95}},\
  \bibinfo {pages} {230801} (\bibinfo {year} {2005})}\BibitemShut {NoStop}%
\bibitem [{\citenamefont {Mitchell}(2017)}]{MitchellQST2017}%
  \BibitemOpen
  \bibfield  {author} {\bibinfo {author} {\bibfnamefont {M.~W.}\ \bibnamefont
  {Mitchell}},\ }\href {https://doi.org/10.1088/2058-9565/aa80c0} {\bibfield
  {journal} {\bibinfo  {journal} {Quantum Science and Technology}\ }\textbf
  {\bibinfo {volume} {2}},\ \bibinfo {pages} {044005} (\bibinfo {year}
  {2017})}\BibitemShut {NoStop}%
\bibitem [{\citenamefont {Lucivero}\ \emph {et~al.}(2016)\citenamefont
  {Lucivero}, \citenamefont {Jim\'enez-Mart\'{\i}nez}, \citenamefont {Kong},\
  and\ \citenamefont {Mitchell}}]{Lucivero2016}%
  \BibitemOpen
  \bibfield  {author} {\bibinfo {author} {\bibfnamefont {V.~G.}\ \bibnamefont
  {Lucivero}}, \bibinfo {author} {\bibfnamefont {R.}~\bibnamefont
  {Jim\'enez-Mart\'{\i}nez}}, \bibinfo {author} {\bibfnamefont
  {J.}~\bibnamefont {Kong}},\ and\ \bibinfo {author} {\bibfnamefont {M.~W.}\
  \bibnamefont {Mitchell}},\ }\href
  {https://doi.org/10.1103/PhysRevA.93.053802} {\bibfield  {journal} {\bibinfo
  {journal} {Phys. Rev. A}\ }\textbf {\bibinfo {volume} {93}},\ \bibinfo
  {pages} {053802} (\bibinfo {year} {2016})}\BibitemShut {NoStop}%
\bibitem [{\citenamefont {Lucivero}\ \emph {et~al.}(2017)\citenamefont
  {Lucivero}, \citenamefont {McDonough}, \citenamefont {Dural},\ and\
  \citenamefont {Romalis}}]{Lucivero2017}%
  \BibitemOpen
  \bibfield  {author} {\bibinfo {author} {\bibfnamefont {V.~G.}\ \bibnamefont
  {Lucivero}}, \bibinfo {author} {\bibfnamefont {N.~D.}\ \bibnamefont
  {McDonough}}, \bibinfo {author} {\bibfnamefont {N.}~\bibnamefont {Dural}},\
  and\ \bibinfo {author} {\bibfnamefont {M.~V.}\ \bibnamefont {Romalis}},\
  }\href {https://doi.org/10.1103/PhysRevA.96.062702} {\bibfield  {journal}
  {\bibinfo  {journal} {Phys. Rev. A}\ }\textbf {\bibinfo {volume} {96}},\
  \bibinfo {pages} {062702} (\bibinfo {year} {2017})}\BibitemShut {NoStop}%
\bibitem [{\citenamefont {Li}\ and\ \citenamefont
  {Novikova}(2022)}]{LiJOSAB2022}%
  \BibitemOpen
  \bibfield  {author} {\bibinfo {author} {\bibfnamefont {J.}~\bibnamefont
  {Li}}\ and\ \bibinfo {author} {\bibfnamefont {I.}~\bibnamefont {Novikova}},\
  }\href {https://doi.org/10.1364/JOSAB.471677} {\bibfield  {journal} {\bibinfo
   {journal} {J. Opt. Soc. Am. B}\ }\textbf {\bibinfo {volume} {39}},\ \bibinfo
  {pages} {2998} (\bibinfo {year} {2022})}\BibitemShut {NoStop}%
\bibitem [{\citenamefont {Horrom}\ \emph {et~al.}(2012)\citenamefont {Horrom},
  \citenamefont {Singh}, \citenamefont {Dowling},\ and\ \citenamefont
  {Mikhailov}}]{HorromPRA2012}%
  \BibitemOpen
  \bibfield  {author} {\bibinfo {author} {\bibfnamefont {T.}~\bibnamefont
  {Horrom}}, \bibinfo {author} {\bibfnamefont {R.}~\bibnamefont {Singh}},
  \bibinfo {author} {\bibfnamefont {J.~P.}\ \bibnamefont {Dowling}},\ and\
  \bibinfo {author} {\bibfnamefont {E.~E.}\ \bibnamefont {Mikhailov}},\ }\href
  {https://doi.org/10.1103/PhysRevA.86.023803} {\bibfield  {journal} {\bibinfo
  {journal} {Phys. Rev. A}\ }\textbf {\bibinfo {volume} {86}},\ \bibinfo
  {pages} {023803} (\bibinfo {year} {2012})}\BibitemShut {NoStop}%
\bibitem [{\citenamefont {Novikova}\ \emph {et~al.}(2015)\citenamefont
  {Novikova}, \citenamefont {Mikhailov},\ and\ \citenamefont
  {Xiao}}]{NovikovaPRA2015}%
  \BibitemOpen
  \bibfield  {author} {\bibinfo {author} {\bibfnamefont {I.}~\bibnamefont
  {Novikova}}, \bibinfo {author} {\bibfnamefont {E.~E.}\ \bibnamefont
  {Mikhailov}},\ and\ \bibinfo {author} {\bibfnamefont {Y.}~\bibnamefont
  {Xiao}},\ }\href {https://doi.org/10.1103/PhysRevA.91.051804} {\bibfield
  {journal} {\bibinfo  {journal} {Phys. Rev. A}\ }\textbf {\bibinfo {volume}
  {91}},\ \bibinfo {pages} {051804} (\bibinfo {year} {2015})}\BibitemShut
  {NoStop}%
\bibitem [{\citenamefont {Troullinou}\ \emph {et~al.}(2021)\citenamefont
  {Troullinou}, \citenamefont {Jim\'enez-Mart\'{\i}nez}, \citenamefont {Kong},
  \citenamefont {Lucivero},\ and\ \citenamefont {Mitchell}}]{Troullinou2021}%
  \BibitemOpen
  \bibfield  {author} {\bibinfo {author} {\bibfnamefont {C.}~\bibnamefont
  {Troullinou}}, \bibinfo {author} {\bibfnamefont {R.}~\bibnamefont
  {Jim\'enez-Mart\'{\i}nez}}, \bibinfo {author} {\bibfnamefont
  {J.}~\bibnamefont {Kong}}, \bibinfo {author} {\bibfnamefont {V.~G.}\
  \bibnamefont {Lucivero}},\ and\ \bibinfo {author} {\bibfnamefont {M.~W.}\
  \bibnamefont {Mitchell}},\ }\href
  {https://doi.org/10.1103/PhysRevLett.127.193601} {\bibfield  {journal}
  {\bibinfo  {journal} {Phys. Rev. Lett.}\ }\textbf {\bibinfo {volume} {127}},\
  \bibinfo {pages} {193601} (\bibinfo {year} {2021})}\BibitemShut {NoStop}%
\bibitem [{\citenamefont {Smullin}\ \emph {et~al.}(2009)\citenamefont
  {Smullin}, \citenamefont {Savukov}, \citenamefont {Vasilakis}, \citenamefont
  {Ghosh},\ and\ \citenamefont {Romalis}}]{Smullin2009}%
  \BibitemOpen
  \bibfield  {author} {\bibinfo {author} {\bibfnamefont {S.~J.}\ \bibnamefont
  {Smullin}}, \bibinfo {author} {\bibfnamefont {I.~M.}\ \bibnamefont
  {Savukov}}, \bibinfo {author} {\bibfnamefont {G.}~\bibnamefont {Vasilakis}},
  \bibinfo {author} {\bibfnamefont {R.~K.}\ \bibnamefont {Ghosh}},\ and\
  \bibinfo {author} {\bibfnamefont {M.~V.}\ \bibnamefont {Romalis}},\ }\href
  {https://doi.org/10.1103/PhysRevA.80.033420} {\bibfield  {journal} {\bibinfo
  {journal} {Phys. Rev. A}\ }\textbf {\bibinfo {volume} {80}},\ \bibinfo
  {pages} {033420} (\bibinfo {year} {2009})}\BibitemShut {NoStop}%
\bibitem [{\citenamefont {Sheng}\ \emph {et~al.}(2013)\citenamefont {Sheng},
  \citenamefont {Li}, \citenamefont {Dural},\ and\ \citenamefont
  {Romalis}}]{Sheng2013}%
  \BibitemOpen
  \bibfield  {author} {\bibinfo {author} {\bibfnamefont {D.}~\bibnamefont
  {Sheng}}, \bibinfo {author} {\bibfnamefont {S.}~\bibnamefont {Li}}, \bibinfo
  {author} {\bibfnamefont {N.}~\bibnamefont {Dural}},\ and\ \bibinfo {author}
  {\bibfnamefont {M.~V.}\ \bibnamefont {Romalis}},\ }\href
  {http://link.aps.org/doi/10.1103/PhysRevLett.110.160802} {\bibfield
  {journal} {\bibinfo  {journal} {Phys. Rev. Lett.}\ }\textbf {\bibinfo
  {volume} {110}},\ \bibinfo {pages} {160802} (\bibinfo {year}
  {2013})}\BibitemShut {NoStop}%
\bibitem [{\citenamefont {Perry}\ \emph {et~al.}(2020)\citenamefont {Perry},
  \citenamefont {Bulatowicz}, \citenamefont {Larsen}, \citenamefont {Walker},\
  and\ \citenamefont {Wyllie}}]{Perry2020}%
  \BibitemOpen
  \bibfield  {author} {\bibinfo {author} {\bibfnamefont {A.~R.}\ \bibnamefont
  {Perry}}, \bibinfo {author} {\bibfnamefont {M.~D.}\ \bibnamefont
  {Bulatowicz}}, \bibinfo {author} {\bibfnamefont {M.}~\bibnamefont {Larsen}},
  \bibinfo {author} {\bibfnamefont {T.~G.}\ \bibnamefont {Walker}},\ and\
  \bibinfo {author} {\bibfnamefont {R.}~\bibnamefont {Wyllie}},\ }\href
  {https://doi.org/10.1364/OE.408486} {\bibfield  {journal} {\bibinfo
  {journal} {Opt. Express}\ }\textbf {\bibinfo {volume} {28}},\ \bibinfo
  {pages} {36696} (\bibinfo {year} {2020})}\BibitemShut {NoStop}%
\bibitem [{\citenamefont {{Gerginov}}\ \emph {et~al.}(2020)\citenamefont
  {{Gerginov}}, \citenamefont {{Pomponio}},\ and\ \citenamefont
  {{Knappe}}}]{Gerginov2020}%
  \BibitemOpen
  \bibfield  {author} {\bibinfo {author} {\bibfnamefont {V.}~\bibnamefont
  {{Gerginov}}}, \bibinfo {author} {\bibfnamefont {M.}~\bibnamefont
  {{Pomponio}}},\ and\ \bibinfo {author} {\bibfnamefont {S.}~\bibnamefont
  {{Knappe}}},\ }\href {https://doi.org/10.1109/JSEN.2020.3002193} {\bibfield
  {journal} {\bibinfo  {journal} {IEEE Sensors Journal}\ }\textbf {\bibinfo
  {volume} {20}},\ \bibinfo {pages} {12684} (\bibinfo {year}
  {2020})}\BibitemShut {NoStop}%
\bibitem [{\citenamefont {Lucivero}\ \emph {et~al.}(2021)\citenamefont
  {Lucivero}, \citenamefont {Lee}, \citenamefont {Dural},\ and\ \citenamefont
  {Romalis}}]{Lucivero2021}%
  \BibitemOpen
  \bibfield  {author} {\bibinfo {author} {\bibfnamefont {V.}~\bibnamefont
  {Lucivero}}, \bibinfo {author} {\bibfnamefont {W.}~\bibnamefont {Lee}},
  \bibinfo {author} {\bibfnamefont {N.}~\bibnamefont {Dural}},\ and\ \bibinfo
  {author} {\bibfnamefont {M.}~\bibnamefont {Romalis}},\ }\href
  {https://link.aps.org/doi/10.1103/PhysRevApplied.15.014004} {\bibfield
  {journal} {\bibinfo  {journal} {Phys. Rev. Applied}\ }\textbf {\bibinfo
  {volume} {15}},\ \bibinfo {pages} {014004} (\bibinfo {year}
  {2021})}\BibitemShut {NoStop}%
\bibitem [{\citenamefont {Bell}\ and\ \citenamefont {Bloom}(1961)}]{BellBloom}%
  \BibitemOpen
  \bibfield  {author} {\bibinfo {author} {\bibfnamefont {W.~E.}\ \bibnamefont
  {Bell}}\ and\ \bibinfo {author} {\bibfnamefont {A.~L.}\ \bibnamefont
  {Bloom}},\ }\href {https://doi.org/10.1103/PhysRevLett.6.280} {\bibfield
  {journal} {\bibinfo  {journal} {Phys. Rev. Lett.}\ }\textbf {\bibinfo
  {volume} {6}},\ \bibinfo {pages} {280} (\bibinfo {year} {1961})}\BibitemShut
  {NoStop}%
\bibitem [{Note1()}]{Note1}%
  \BibitemOpen
  \bibinfo {note} {The number density $n$ is adjusted by stabilizing the
  temperature $T$ of the ceramic oven enclosing the vapor cell. For \protect
  \textsuperscript {87}Rb and $T>\SI {312.5}{\kelvin }$ we compute $n$ in \SI
  {}{atoms\per \centi \meter \cubed } using $\protect \qopname \relax
  o{log}_{10}(Tn) = 21.866 + A -B/T$ with coefficients $A = 4.312$ and $B=\SI
  {4040}{K}$ \cite {KongNC2020}.}\BibitemShut {Stop}%
\bibitem [{\citenamefont {Predojevi\ifmmode~\acute{c}\else \'{c}\fi{}}\ \emph
  {et~al.}(2008)\citenamefont {Predojevi\ifmmode~\acute{c}\else \'{c}\fi{}},
  \citenamefont {Zhai}, \citenamefont {Caballero},\ and\ \citenamefont
  {Mitchell}}]{Predojevic2008}%
  \BibitemOpen
  \bibfield  {author} {\bibinfo {author} {\bibfnamefont {A.}~\bibnamefont
  {Predojevi\ifmmode~\acute{c}\else \'{c}\fi{}}}, \bibinfo {author}
  {\bibfnamefont {Z.}~\bibnamefont {Zhai}}, \bibinfo {author} {\bibfnamefont
  {J.~M.}\ \bibnamefont {Caballero}},\ and\ \bibinfo {author} {\bibfnamefont
  {M.~W.}\ \bibnamefont {Mitchell}},\ }\href
  {https://link.aps.org/doi/10.1103/PhysRevA.78.063820} {\bibfield  {journal}
  {\bibinfo  {journal} {Phys. Rev. A}\ }\textbf {\bibinfo {volume} {78}},\
  \bibinfo {pages} {063820} (\bibinfo {year} {2008})}\BibitemShut {NoStop}%
\bibitem [{\citenamefont {Savukov}\ and\ \citenamefont
  {Romalis}(2005)}]{Savukov2005}%
  \BibitemOpen
  \bibfield  {author} {\bibinfo {author} {\bibfnamefont {I.~M.}\ \bibnamefont
  {Savukov}}\ and\ \bibinfo {author} {\bibfnamefont {M.~V.}\ \bibnamefont
  {Romalis}},\ }\href {https://doi.org/10.1103/PhysRevA.71.023405} {\bibfield
  {journal} {\bibinfo  {journal} {Phys. Rev. A}\ }\textbf {\bibinfo {volume}
  {71}},\ \bibinfo {pages} {023405} (\bibinfo {year} {2005})}\BibitemShut
  {NoStop}%
\bibitem [{Sup()}]{SupplementalMaterial}%
  \BibitemOpen
  \href@noop {} {}\bibinfo {note} {Supplemental online material, URL to be
  inserted by journal.}\BibitemShut {Stop}%
\bibitem [{\citenamefont {Gerginov}\ \emph {et~al.}(2017)\citenamefont
  {Gerginov}, \citenamefont {Krzyzewski},\ and\ \citenamefont
  {Knappe}}]{Gerginov2017}%
  \BibitemOpen
  \bibfield  {author} {\bibinfo {author} {\bibfnamefont {V.}~\bibnamefont
  {Gerginov}}, \bibinfo {author} {\bibfnamefont {S.}~\bibnamefont
  {Krzyzewski}},\ and\ \bibinfo {author} {\bibfnamefont {S.}~\bibnamefont
  {Knappe}},\ }\href {https://doi.org/10.1364/JOSAB.34.001429} {\bibfield
  {journal} {\bibinfo  {journal} {J. Opt. Soc. Am. B}\ }\textbf {\bibinfo
  {volume} {34}},\ \bibinfo {pages} {1429} (\bibinfo {year}
  {2017})}\BibitemShut {NoStop}%
\bibitem [{\citenamefont {Han}\ \emph {et~al.}(2016)\citenamefont {Han},
  \citenamefont {Wen}, \citenamefont {He}, \citenamefont {Yang}, \citenamefont
  {Wang},\ and\ \citenamefont {Wang}}]{HanOE2016}%
  \BibitemOpen
  \bibfield  {author} {\bibinfo {author} {\bibfnamefont {Y.}~\bibnamefont
  {Han}}, \bibinfo {author} {\bibfnamefont {X.}~\bibnamefont {Wen}}, \bibinfo
  {author} {\bibfnamefont {J.}~\bibnamefont {He}}, \bibinfo {author}
  {\bibfnamefont {B.}~\bibnamefont {Yang}}, \bibinfo {author} {\bibfnamefont
  {Y.}~\bibnamefont {Wang}},\ and\ \bibinfo {author} {\bibfnamefont
  {J.}~\bibnamefont {Wang}},\ }\href {https://doi.org/10.1364/OE.24.002350}
  {\bibfield  {journal} {\bibinfo  {journal} {Optics Express}\ }\textbf
  {\bibinfo {volume} {24}},\ \bibinfo {pages} {2350} (\bibinfo {year}
  {2016})}\BibitemShut {NoStop}%
\bibitem [{\citenamefont {Stockton}(2007)}]{StocktonThesis2007}%
  \BibitemOpen
  \bibfield  {author} {\bibinfo {author} {\bibfnamefont {J.~K.}\ \bibnamefont
  {Stockton}},\ }\emph {\bibinfo {title} {Continuous quantum measurement of
  cold alkali-atom spins.}},\ \href
  {http://resolver.caltech.edu/CaltechETD:etd-02172007-172548} {Ph.D. thesis},\
  \bibinfo  {school} {California Institute of Technology} (\bibinfo {year}
  {2007})\BibitemShut {NoStop}%
\bibitem [{\citenamefont {Appelt}\ \emph {et~al.}(1999)\citenamefont {Appelt},
  \citenamefont {Ben-Amar~Baranga}, \citenamefont {Young},\ and\ \citenamefont
  {Happer}}]{Appelt1999}%
  \BibitemOpen
  \bibfield  {author} {\bibinfo {author} {\bibfnamefont {S.}~\bibnamefont
  {Appelt}}, \bibinfo {author} {\bibfnamefont {A.}~\bibnamefont
  {Ben-Amar~Baranga}}, \bibinfo {author} {\bibfnamefont {A.~R.}\ \bibnamefont
  {Young}},\ and\ \bibinfo {author} {\bibfnamefont {W.}~\bibnamefont
  {Happer}},\ }\href {https://doi.org/10.1103/PhysRevA.59.2078} {\bibfield
  {journal} {\bibinfo  {journal} {Phys. Rev. A}\ }\textbf {\bibinfo {volume}
  {59}},\ \bibinfo {pages} {2078} (\bibinfo {year} {1999})}\BibitemShut
  {NoStop}%
\bibitem [{\citenamefont {Ito}\ \emph {et~al.}(2016)\citenamefont {Ito},
  \citenamefont {Sato}, \citenamefont {Kamada},\ and\ \citenamefont
  {Kobayashi}}]{ItoKobayashi2016}%
  \BibitemOpen
  \bibfield  {author} {\bibinfo {author} {\bibfnamefont {Y.}~\bibnamefont
  {Ito}}, \bibinfo {author} {\bibfnamefont {D.}~\bibnamefont {Sato}}, \bibinfo
  {author} {\bibfnamefont {K.}~\bibnamefont {Kamada}},\ and\ \bibinfo {author}
  {\bibfnamefont {T.}~\bibnamefont {Kobayashi}},\ }\href
  {https://doi.org/10.1364/OE.24.015391} {\bibfield  {journal} {\bibinfo
  {journal} {Opt. Express}\ }\textbf {\bibinfo {volume} {24}},\ \bibinfo
  {pages} {15391} (\bibinfo {year} {2016})}\BibitemShut {NoStop}%
\bibitem [{\citenamefont {Colangelo}\ \emph {et~al.}(2017)\citenamefont
  {Colangelo}, \citenamefont {Ciurana}, \citenamefont {Bianchet}, \citenamefont
  {Sewell},\ and\ \citenamefont {Mitchell}}]{ColangeloN2017}%
  \BibitemOpen
  \bibfield  {author} {\bibinfo {author} {\bibfnamefont {G.}~\bibnamefont
  {Colangelo}}, \bibinfo {author} {\bibfnamefont {F.~M.}\ \bibnamefont
  {Ciurana}}, \bibinfo {author} {\bibfnamefont {L.~C.}\ \bibnamefont
  {Bianchet}}, \bibinfo {author} {\bibfnamefont {R.~J.}\ \bibnamefont
  {Sewell}},\ and\ \bibinfo {author} {\bibfnamefont {M.~W.}\ \bibnamefont
  {Mitchell}},\ }\href {https://doi.org/10.1038/nature21434} {\bibfield
  {journal} {\bibinfo  {journal} {Nature}\ }\textbf {\bibinfo {volume} {543}},\
  \bibinfo {pages} {525} (\bibinfo {year} {2017})}\BibitemShut {NoStop}%
\bibitem [{\citenamefont {Limes}\ \emph {et~al.}(2020)\citenamefont {Limes},
  \citenamefont {Foley}, \citenamefont {Kornack}, \citenamefont {Caliga},
  \citenamefont {McBride}, \citenamefont {Braun}, \citenamefont {Lee},
  \citenamefont {Lucivero},\ and\ \citenamefont {Romalis}}]{Limes2020}%
  \BibitemOpen
  \bibfield  {author} {\bibinfo {author} {\bibfnamefont {M.}~\bibnamefont
  {Limes}}, \bibinfo {author} {\bibfnamefont {E.}~\bibnamefont {Foley}},
  \bibinfo {author} {\bibfnamefont {T.}~\bibnamefont {Kornack}}, \bibinfo
  {author} {\bibfnamefont {S.}~\bibnamefont {Caliga}}, \bibinfo {author}
  {\bibfnamefont {S.}~\bibnamefont {McBride}}, \bibinfo {author} {\bibfnamefont
  {A.}~\bibnamefont {Braun}}, \bibinfo {author} {\bibfnamefont
  {W.}~\bibnamefont {Lee}}, \bibinfo {author} {\bibfnamefont {V.}~\bibnamefont
  {Lucivero}},\ and\ \bibinfo {author} {\bibfnamefont {M.}~\bibnamefont
  {Romalis}},\ }\href {https://doi.org/10.1103/PhysRevApplied.14.011002}
  {\bibfield  {journal} {\bibinfo  {journal} {Phys. Rev. Applied}\ }\textbf
  {\bibinfo {volume} {14}},\ \bibinfo {pages} {011002} (\bibinfo {year}
  {2020})}\BibitemShut {NoStop}%
\bibitem [{\citenamefont {Limes}\ \emph {et~al.}(2018)\citenamefont {Limes},
  \citenamefont {Sheng},\ and\ \citenamefont {Romalis}}]{Limes2018}%
  \BibitemOpen
  \bibfield  {author} {\bibinfo {author} {\bibfnamefont {M.~E.}\ \bibnamefont
  {Limes}}, \bibinfo {author} {\bibfnamefont {D.}~\bibnamefont {Sheng}},\ and\
  \bibinfo {author} {\bibfnamefont {M.~V.}\ \bibnamefont {Romalis}},\ }\href
  {https://doi.org/10.1103/PhysRevLett.120.033401} {\bibfield  {journal}
  {\bibinfo  {journal} {Phys. Rev. Lett.}\ }\textbf {\bibinfo {volume} {120}},\
  \bibinfo {pages} {033401} (\bibinfo {year} {2018})}\BibitemShut {NoStop}%
\bibitem [{\citenamefont {Zieli\ifmmode~\acute{n}\else \'{n}\fi{}ska}\ \emph
  {et~al.}(2023)\citenamefont {Zieli\ifmmode~\acute{n}\else \'{n}\fi{}ska},
  \citenamefont {van~der Laan}, \citenamefont {Norrman}, \citenamefont
  {Rimlinger}, \citenamefont {Reimann}, \citenamefont {Novotny},\ and\
  \citenamefont {Frimmer}}]{ZielinskaPRL2023b}%
  \BibitemOpen
  \bibfield  {author} {\bibinfo {author} {\bibfnamefont {J.~A.}\ \bibnamefont
  {Zieli\ifmmode~\acute{n}\else \'{n}\fi{}ska}}, \bibinfo {author}
  {\bibfnamefont {F.}~\bibnamefont {van~der Laan}}, \bibinfo {author}
  {\bibfnamefont {A.}~\bibnamefont {Norrman}}, \bibinfo {author} {\bibfnamefont
  {M.}~\bibnamefont {Rimlinger}}, \bibinfo {author} {\bibfnamefont
  {R.}~\bibnamefont {Reimann}}, \bibinfo {author} {\bibfnamefont
  {L.}~\bibnamefont {Novotny}},\ and\ \bibinfo {author} {\bibfnamefont
  {M.}~\bibnamefont {Frimmer}},\ }\href
  {https://doi.org/10.1103/PhysRevLett.130.203603} {\bibfield  {journal}
  {\bibinfo  {journal} {Phys. Rev. Lett.}\ }\textbf {\bibinfo {volume} {130}},\
  \bibinfo {pages} {033401} (\bibinfo {year} {2023})}\BibitemShut {NoStop}%
\bibitem [{\citenamefont {Kong}\ \emph {et~al.}(2020)\citenamefont {Kong},
  \citenamefont {Jim{\'e}nez-Mart{\'\i}nez}, \citenamefont {Troullinou},
  \citenamefont {Lucivero}, \citenamefont {T{\'o}th},\ and\ \citenamefont
  {Mitchell}}]{KongNC2020}%
  \BibitemOpen
  \bibfield  {author} {\bibinfo {author} {\bibfnamefont {J.}~\bibnamefont
  {Kong}}, \bibinfo {author} {\bibfnamefont {R.}~\bibnamefont
  {Jim{\'e}nez-Mart{\'\i}nez}}, \bibinfo {author} {\bibfnamefont
  {C.}~\bibnamefont {Troullinou}}, \bibinfo {author} {\bibfnamefont {V.~G.}\
  \bibnamefont {Lucivero}}, \bibinfo {author} {\bibfnamefont {G.}~\bibnamefont
  {T{\'o}th}},\ and\ \bibinfo {author} {\bibfnamefont {M.~W.}\ \bibnamefont
  {Mitchell}},\ }\href {https://doi.org/10.1038/s41467-020-15899-1} {\bibfield
  {journal} {\bibinfo  {journal} {Nature Communications}\ }\textbf {\bibinfo
  {volume} {11}},\ \bibinfo {pages} {2415} (\bibinfo {year}
  {2020})}\BibitemShut {NoStop}%
\bibitem [{Note2()}]{Note2}%
  \BibitemOpen
  \bibinfo {note} {The squeezing bandwidth in the experiment is $\approx \SI
  {8}{\mega \hertz }$ \cite {Predojevic2008}, so ${\protect \cal S}_{N_{S_2}}$
  is to a good approximation constant over the $\sim \SI {}{\kilo \hertz }$
  bandwidth of interest.}\BibitemShut {Stop}%
\bibitem [{\citenamefont {Troullinou}(2022)}]{Troullinou_2022}%
  \BibitemOpen
  \bibfield  {author} {\bibinfo {author} {\bibfnamefont {C.}~\bibnamefont
  {Troullinou}},\ }\href {http://hdl.handle.net/2117/379458} {Ph.D. thesis},\
  \bibinfo  {school} {UPC, Institut de Ciències Fotòniques} (\bibinfo {year}
  {2022})\BibitemShut {NoStop}%
\end{thebibliography}%

\newpage

\begin{widetext}
\newpage
\noindent
{\large Supplemental material for:\textit{\thetitle}}
\end{widetext} 

\setcounter{page}{1}
\renewcommand{\thepage}{S\arabic{page}} 
\renewcommand{\thesection}{S\arabic{section}}  
\renewcommand{\thetable}{S\arabic{table}}  
\renewcommand{\thefigure}{S\arabic{figure}}
\renewcommand{\theequation}{S\arabic{equation}} 
\setcounter{figure}{0} 

\begin{figure}[t]
    \centering
    \includegraphics[width=0.8 \columnwidth]{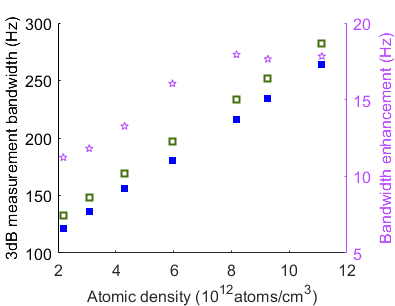}
    \caption{\textbf{Density dependence of \SI{3}{\decibel} measurement bandwidth.}  Measurement bandwidth for coherent (filled blue) and squeezed light (hollow green) probing (left) and quantum enhancement of \SI{3}{\decibel} measurement bandwidth $\omega_{\SI{3}{\decibel}}|_{\mathrm{{Coh}}} - \omega_{\SI{3}{\decibel}}|_{\mathrm{{Sq}}} $ (stars, right).  
    } 
    \label{fig:w3dBVSdensity}
\end{figure}

\section{Quantum advantage in measurement bandwidth} Squeezed-light probing suppresses the photon shot noise level, increasing the ratio $\Sat /\Sph$ and therefore improves the \SI{3}{\decibel} measurement bandwidth 
$\omega_{\SI{3}{\decibel}} \equiv \Delta \omega \sqrt{ \Sat /\Sph +1}$ \cite{Troullinou2021}. 
As shown in Fig.~\ref{fig:w3dBVSdensity},
$\omega_{\SI{3}{\decibel}}$ increases when squeezed light is used for probing, at all measured densities.

\section{Spin projection noise extrapolation to $n = \SI{0.5e12}{atoms\per\centi\meter\cubed}$.}
\begin{figure}[t]
    \centering
    \includegraphics[width=0.8 \columnwidth]{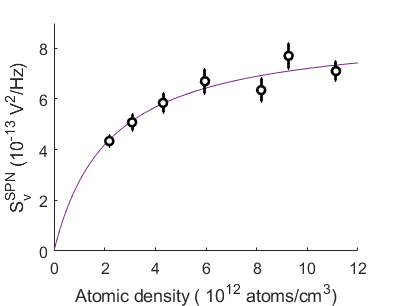}
    \caption{\textbf{Density dependence of spin projection noise level.}  
    Points show $\Sat$ from a fit of Eq.~\ref{eq:SvInTermsOfSNAndSsigma} to measured spin noise spectra, as in Fig.~\ref{fig:SNSandMAGspectra95C}(a), for various atomic number densities $n$. Violet line shows fit with $\alpha n/(\beta +n)$, with $\alpha$ and $\beta$ free parameters. From this fit, $\Sat= \SI{1.6e-13}{\volt\squared\per\hertz}$ at $n = \SI{0.5e12}{atoms\per\centi\meter\cubed}$.}
    \label{fig:SatVSdensity}
\end{figure}

We compute the frequency dependent spin projection noise $\Sat \cal{L}(\omega)$ (grey dash-dotted curve in Fig.~\ref{fig:DoubleSmiles}) with $\Delta\omega$ found by extrapolation of the linear fit of the data in Fig.~\ref{fig:experimentalSetup} (c) (top) to $n = \SI{0.5e12}{atoms\per\centi\meter\cubed}$. To estimate the $\Sat$ for $n = \SI{0.5e12}{atoms\per\centi\meter\cubed}$, we fit the spin noise spectra measured for each density to Eq.~\ref{eq:SvInTermsOfSNAndSsigma}. Then we fit the values $\Sat(n)$ to the function $An/(B+n)$, with A and B free parameters and extrapolate to $n = \SI{0.5e12}{atoms\per\centi\meter\cubed} $, as seen in Fig.~\ref{fig:SatVSdensity}.


\section{Derivation of Eq. (2)}

Under the simplifying assumption of a single hyperfine state, and with a linearized quantum noise analysis, Troullinou et al.~\cite[]{Troullinou2021}, derived the quantum-noise-limited sensitivity of the Bell-Bloom (BB) optically-pumped magnetometer (OPM). We start with the results of that analysis, in the form of Eqs. (S42-S46) from that work. 

From \cite[Eqs. (S42-S43)]{Troullinou2021}, the square responsivity (ratio of optical to magnetic signal power) is
\bea
\label{eq:S2SignalQSlopeSqequiv}
|R(\omega)|^2 & = &  \gamma^2 \frac{\langle u \rangle^2}{\Delta \omega^2} {\cal L}(\omega).
\eea
 where $\gamma$ is the  bare \textsuperscript{87}Rb gyromagnetic ratio, $\langle u \rangle$ is the signal amplitude, $\Delta\omega$ is the HWHM  magnetic resonance linewidth, and $\mathcal{L}(\omega) \equiv {\Delta\omega^2}/(\omega^2 + \Delta\omega^2)$ is the normalized square responsivity.

From \cite[Eq. (S45-S46)]{Troullinou2021}, the magnetometer sensitivity (equivalent magnetic noise spectrum) is 
\begin{equation}
{\cal S}_B(\omega) = |R(\omega)|^{-2} {\cal S}_v(\omega),
\end{equation}
where the optical noise spectrum is
\be
\label{eq:SMSv}
{\cal S}_v(\omega) = {\cal S}_{N_{S_2}} + {\cal L} (\omega){\cal S}_\sigma,
\ee
${\cal S}_{N_{S_2}} \equiv \Sph$ is the noise in the $S_2$ Stokes component (the PSN), and ${\cal S}_\sigma \equiv \Sat$ is the SPN contribution to $v$. These noises are assumed white, and thus without $\omega$ dependence \footnote{The squeezing bandwidth in the experiment is $\approx \SI{8}{\mega\hertz}$ \cite{Predojevic2008}, so ${\cal S}_{N_{S_2}}$ is to a good approximation constant over the $\sim\SI{}{\kilo\hertz}$ bandwidth of interest.}. In Eq.~\ref{eq:SMSv}, MBA is absent due to the back-action evading nature of the BB OPM with Faraday rotation readout. 

For greater generality, we include MBA as follows: we assume white quantum noise in probe optical variables such as the Stokes parameter $S_3(t)$, and note that the linearized response of the spin observables to this additional perturbation will have noise power $\propto \mathcal{L}(\omega)$, just as does the white noise associated with spin relaxation (the second term in Eq.~\ref{eq:SMSv}) \cite{Troullinou2021}. We can thus generically add a MBA term ${\cal L} (\omega){\cal S}_\mathrm{MBA}$ to Eq.~\ref{eq:SMSv}). We thus have the optical noise 
\be
\label{eq:SMSvWithMBA}
{\cal S}_v(\omega) = \Sph + {\cal L} (\omega)\Sat + \mathcal{L} (\omega)\Smba
\ee
and the equivalent magnetic noise spectrum  
\bea
\label{eq:SBFromSM}
\mathcal{S}_B(\omega) & = & \SBph + \SBat + \SBmba \nonumber \\ 
 & = &\frac{\Delta \omega^2}{\gamma^2\langle u \rangle^2 } [ \frac{\Sph}{\mathcal{L}(\omega)} + \Sat + \Smba].
\eea

We now discuss how each of these contributions scales with the atomic number density $n$ and the photon fluxes $\Phi_\mathrm{pr}$ and $\Phi_\mathrm{pump}$ of the probe and pump beams, respectively. 

\subsection{Spin polarization}
From \cite[Eq. (S35)]{Troullinou2021}, the zero-th order spin polarization is
\bea
\label{eq:ZeroOrderSteadyRot3}
\mathbf{F}\supzero  & \propto &  n\frac{\Phi_\mathrm{pump}}{\Delta\omega},
\eea
where $\Delta \omega = \Gamma + \Gamma_2 n + \bar{P}$, $\Gamma_2 n$ is the collisional broadening, $\bar{P} \propto \Phi_\mathrm{pump}$ is the broadening due to pump light, and $\Gamma$ is independent of $n, \Phi_\mathrm{pump}$. The resulting saturation of  $\mathbf{F}\supzero$ with $n$ can be seen in Fig.~\ref{fig:experimentalSetup} (c)(top).

\subsection{Signal Amplitude}
The signal amplitude is \cite{Troullinou2021} 
\begin{equation}
\label{eq:vBarWithG}
\langle u(\omega)\rangle = G S_1 \rho\supzero ,
\end{equation}
where $G$ is the light-atom coupling, $S_1 \propto \Phi_\mathrm{pr}$ is the input polarization, 
and the in-phase component of ${F}_z\supzero$ is 
\begin{equation} 
\rho\supzero
\propto \mathbf{F}\supzero \propto n\frac{\Phi_\mathrm{pump}}{\Delta\omega}.
\end{equation}  
We thus have
\begin{equation}
\langle u(\omega)\rangle \propto n \frac{\Phi_\mathrm{pr} \Phi_\mathrm{pump}}{\Delta \omega}.
\end{equation}

\subsection{Spin projection noise}

From \cite[Eq.~(S44)]{Troullinou2021},  the spin projection noise contribution to the signal $v$ is  
\begin{equation}
\Sat \equiv \mathcal{S}_\sigma \equiv \frac{G^2 S_1^2}{\Delta\omega^2} \mathcal{S}_{\mathcal{I}[N_{F_+}]}(\omega), 
\label{eq:Sat}
\end{equation}
where G the atom-light coupling, $S_1 \propto \Phi_\mathrm{pr}$ is the mean value of the Stokes parameter and $\mathcal{S}_{\mathcal{I}[N_{F_+}]}(\omega)$ is the Fourier transform of the cross correlation function
\bea
\label{eq:NAutocorr}
\langle N_{F_i}(t)N_{F_j}(t') \rangle &=& 2 \frac{f(f+1)}{3} N_A \Delta\omega \delta_{ij} \delta(t-t'),
\hspace{9mm} 
\eea
where $f$ is the spin quantum number and $N_A$ is the effective atom number in the probe beam. Evidently $
\langle N_{F_i}(t)N_{F_j}(t') \rangle \propto n \Delta\omega$ 
and thus the spin projection noise contribution to the signal scales as 
 \bea
\Sat \propto \frac{\Phi_\mathrm{pr}^2}{\Delta\omega} n.
 \eea 

\subsection{Photon shot noise}
Similarly, the optical quantum noise in the Stokes components $\mathcal{S}_{N_{S_j}}$, $j \in \{2,3\}$ is the Fourier transform of the autocorrelation function $\langle S_j(t) S_j(t')  \rangle$. 
Including the squeezing factor $\xi^2$ these are 
\bea
\langle S_2(t) S_2(t') \rangle &=& \frac{1}{2} \xi^2 \langle S_1(t) \rangle\, \delta(t-t') 
\\
\langle S_3(t) S_3(t') \rangle &=& \frac{1}{2} \xi^{-2} \langle S_1(t) \rangle\, \delta(t-t') 
\eea
c.f. \cite[Eq.~(2.45)]{Troullinou_2022}, such that 
\bea
\mathcal{S}_{N_{S_2}} &\propto& \xi^2 \Phi_\mathrm{pr} 
\\
\mathcal{S}_{N_{S_3}} &\propto& \xi^{-2} \Phi_\mathrm{pr},
\eea   
and thus 
\bea
\Sph \equiv \mathcal{S}_{N_{S_2}} \propto \xi^2\Phi_\mathrm{pr}.
\label{eq:Sph}
\eea

\subsection{Measurement back-action noise}

As in Fig.~\ref{fig:experimentalSetup}, in \cite{Troullinou2021} the B-field is along the $\hat{x}$ direction, and the probing is along $\hat{z}$.  Then, \cite[Eq. (S15)]{Troullinou2021} describes how the Stokes component $S_3(t)$, with quantum noise as described immediately above, produces ac Stark shifts that introduce a noise term $GS_3(t)\hat{z}\times \bF$ at first order:
\begin{equation} 
\frac{d}{dt} \bF\supone = - \gamma \bB\supone  \times \bF\supzero + GS_3(t)\hat{z}\times \bF\supzero + [\mathrm{other~terms}].
\end{equation} 
The signal term $-\gamma \bB\supone  \times \bF\supzero$ produces a signal $v$ proportional to $\bB\supone_x$.  The noise term has the same effect as a white-noise $B$-field along $\hat{z}$. For the scenario considered in \cite{Troullinou2021}, with a B-field orthogonal to the probing direction, the lowest-order effect of this noise term appears only in the component of $\mathbf{F}$ parallel to $\mathbf{B}$, and thus does not affect the signal quadratures $u$ or $v$, which record the evolution of the components of $\mathbf{F}$ perpendicular to $\bB$.  For this reason, the MBA term in \cite{Troullinou2021} vanishes. 

In other scenarios, for example if the probe beam along direction  $\hat{\zeta}$ makes an angle $\theta$ with the B-field, the resulting $GS_3(t)\hat{\zeta}\times \bF\supzero$ term will have an $\hat{x}$ component $\propto \cos \theta$, which contributes to the signal alongside $B\supone_x$. After cycle-averaging,  
\begin{equation} 
\frac{d}{dt} \bF\supone = [- \gamma B_x\supone + GS_3(t)\cos\theta] \hat{x}\times \bF\supzero + [\mathrm{other~terms}].
\end{equation} 
Noting that the magnetic signal $B_x\supone $ and noise $S_3$ enter in exactly the same way apart from multiplicative factors, and using the $B$-field contribution from Eq.~\ref{eq:S2SignalQSlopeSqequiv}, the MBA noise contribution is 
\bea
\Smba & \propto & \frac{\langle u \rangle^2 G^2}{\Delta \omega^2} \langle S_3(t)S_3(t') \rangle
\nonumber \\ 
& \propto & \frac{\Phi_\mathrm{pr}^2}{\Delta\omega^2}  \frac{n^2 \Phi_\mathrm{pump}^2}{\Delta\omega^2} \xi^{-2}\Phi_\mathrm{pr}.
\label{eq:SmbaMWM}
\eea

\subsection{OPM power spectral density}
Substituting \ref{eq:Sat},\ref{eq:Sph} and \ref{eq:SmbaMWM} into \ref{eq:SBFromSM} we find 
\begin{eqnarray}
\label{eq:SBAgainWithMBA}
 \mathcal{S}_B(\omega) 
&\propto& \frac{\Delta \omega^3  }{n \Phi_\mathrm{pump}^2 }
+ 
X \xi^2 \frac{\Delta \omega^2  (\omega^2 + \Delta\omega^2)}{n^2  \Phi_\mathrm{pump}^2 \Phi_\mathrm{pr} }
+
Y \frac{ \Phi_\mathrm{pr}}{\xi^2}
 , \hspace{9mm}
\end{eqnarray}
where $X$ and $Y$ are constants that depend on the coupling factor (and thus the probe detuning) and degree of back-action evasion, respectively, but are independent of $n$, $\Phi_\mathrm{pump}$ and $\Phi_\mathrm{pr}$.

\end{document}